# Measuring Physical Activity of Older Adults via Smartwatch and Stigmergic Receptive Fields.

Antonio L. Alfeo, Mario G.C.A. Cimino and Gigliola Vaglini
*Department of Information Engineering, Università di Pisa, largo Lazzarino 1, Pisa, Italy*
*luca.alfeo@ing.unipi.it, {mario.cimino, gigliola.vaglini}@unipi.it.*

Abstract: Physical activity level (PAL) in older adults can enhance healthy aging, improve functional capacity, and prevent diseases. It is known that human annotations of PAL can be affected by subjectivity and inaccuracy. Recently developed smart devices can allow a non-invasive, analytic, and continuous gathering of physiological signals. We present an innovative computational system fed by signals of heartbeat rate, wrist motion and pedometer sensed by a smartwatch. More specifically, samples of each signal are aggregated by functional structures called *trails*. The trailing process is inspired by *stigmergy*, an insects' coordination mechanism, and is managed by computational units called *stigmergic receptive fields* (SRFs). SRFs, which compute the similarity between trails, are arranged in a *stigmergic perceptron* to detect a collection of micro-behaviours of the raw signal, called *archetypes*. A SRF is adaptive to subjects: its structural parameters are tuned by a differential evolution algorithm. SRFs are used in a multilayer architecture, providing further levels of processing to realize macro analyses in the application domain. As a result, the architecture provides a daily PAL, useful to detect behavioural shift indicating initial signs of disease or deviations in performance. As a proof of concept, the approach has been experimented on three subjects.

## 1 INTRODUCTION AND MOTIVATION

Resistance and physiological reserves decrease in older people, resulting in a risk of adverse health effects. This state of vulnerability is called *frailty* (Fontecha, 2011) and is assessed taking into account the physical activity level (PAL), among other factors (Fontecha, 2013). Nowadays, physicians detect frailty by means of specialized questionnaires and physical tests performed in dedicated facilities. However, the number of pre-frail elder people, which identifies a high risk of progressing to frailty, is increasing beyond the facilities potential. On the other hand, human-driven test scores may be insufficient and inaccurate for detecting physical habits (Boletsis, 2015), and can be affected by certain degree of subjectivity (Jansen, 2015).

Today the great availability of general purpose wearable devices offers a new opportunity for non-invasive healthcare monitoring. Some watch-like systems have been already developed to monitor specific user's physical activities, exploiting heart rate and motion signals. Actually, much work has to be done before such systems can be regularly managed: the detection of a specific physical activity usually implies complex techniques, including machine learning and probabilistic modelling. For a widespread adoption the system should be highly flexible, handle uncertainty, and allow a personalization of what to monitor and how to notice it. In this paper we propose to use a smartwatch to detect the physical activity level rather than a specific physical activity. This approach can provide enough benefits to warrant widespread adoption. For this purpose, we studied a suitable computational architecture with adaptive setting and configuration. In the proposed architecture, the input samples are managed by computational units called *Stigmergic Receptive Fields* (SRFs), organized into a multilayer connectionist architecture (Cimino, 2009), and adapted to contextual behavior by means of the Differential Evolution algorithm.

The paper is structured as follows. Section 2 discusses the research works dealing with the use of smartwatch for activity monitoring. In Section 3, an ontological and architectural view of our system is presented. Section 4 covers the experimental studies on three case studies. Finally, Section 5 summarizes conclusions and future work.

## 2 RELATED WORK

In the literature, some studies have recently proved that is possible to distinguish amongst different human activities, as well as to measure the physical effort, through wearable device and data-driven classification techniques (Abbate, 2012).

In (Bonomi, 2010) 30 healthy subjects have been monitored for 14 days, using: (i) a triaxial accelerometer for movement registration to calculate the activity counts per day; (ii) a laboratory equipment (indirect calorimetry) to calculate the total energy expenditure in free living conditions; (iii) a respiration chamber to measure during an overnight stay the sleeping metabolic rate. The activity energy expenditure and the physical activity level are determined from total energy expenditure and sleeping metabolic rate. A direct linear association was observed between the activity counts per day and the physical activity level. A multiple-linear regression model predicted 76% of the variance in total energy expenditure, which is a very high accuracy for predicting free-living energy expenditure.

Guiry *et al.* (2014) gathered samples from 10 subjects, each equipped with a smartphone and a smartwatch, exploiting all available sensors (tri-axial accelerometer, tri-axial magnetometer, tri-axial gyroscope, GPS, light and pressure sensors). Subjects were asked to perform specific physical activities during three different gathering phases. In the proposed approach, data samples are first pre-processed via Principal Component Analysis. Subsequently, the data set is used to classify the physical activities, by using five well-known learning algorithms: C4.5, CART, Naïve Bayes, Multi-Layer Perceptrons and Support Vector Machines. Results show that the system correctly classifies the activities with a percentage of 95% when using a smartphone and 89% when using a smartwatch.

Parkka *et al.* (2007) estimate the intensity of physical activity attaching accelerometers and gyroscopes to ankle, wrist and hip. The results are compared to metabolic equivalent measures obtained by means of a portable system used for testing cardiopulmonary exercise. Experiments are made with 11 subjects carrying out everyday tasks, including ironing, vacuuming, walking, running, and cycling on exercise bicycle (ergometer). The authors have calculated a linear correlation between accelerometers signals and metabolic equivalent up to 0.86.

Zhu *et al.* (2015) estimate physical activities energy expenditure using wearable devices in different activities: walking, standing, climbing upstairs or downstairs. More specifically, a Convolution Neural Networks is used to automatically detect important features from data collected from triaxial accelerometer and heart rate sensors. The results are compared with the state-of-the-art of linear regression and artificial neural networks applied to specific activities, obtaining a mean square error of 1.12 which is about 35% lower than existing models.

In this paper we apply two strategies for improving the state of the art.

*a) Computational strategy*. The bio-inspired paradigm of emergent systems (e.g. manifested by societies of insects) is exploited for spatio-temporal data granulation. With this paradigm, the single data sample embodies a domain-agnostic micro-behavior, interacting with other samples. The principles of connectionism are also applied to achieve new levels of abstraction without explicit knowledge encoded (Barsocchi, 2015). The purpose is to enable the production of macro-behavior phenomena as an emergent process of evolution of interconnected processing units.

*b) Application strategy*. The purpose is to generate continuous behavioral data through general-purpose and non-intrusive devices. To detect behavioral patterns used in broad-spectrum assessment: behavior shift to discover initial signs of disease or deviations in performance.

Thus, the detection of explicit user activities and diagnosis of specific diseases are not within the scope of our approach. The next section presents an ontological view of the approach and a core set of functionalities.

## 3 CORE CONCEPTS AND FUNCTIONAL DESIGN

This section unfolds the core concepts and their relationships in an ontological view. In Fig. 1 base concepts are enclosed in grey ovals and connected by properties (represented as black directed edges), whereas specialized concepts are enclosed in white ovals and connected to base concepts by the *is-a* property (represented as white directed edge). More specifically, *an older adult performs a physical activity*, which *is measured by many behavior patterns*. The *older adult wears a smart watch*, which *gathers the physiological signals: pedometer, wrist motion,* and *heartbeat rate*. A *reference signal*

*is a specific kind of physiological signal*, an *archetype is a special kind of reference signal*. A *physiological signal releases a sequence of marks*, which *aggregates in trails. Evolution adapts mark and trail. Similarity compares two trails*, and *detects a behavior pattern*.

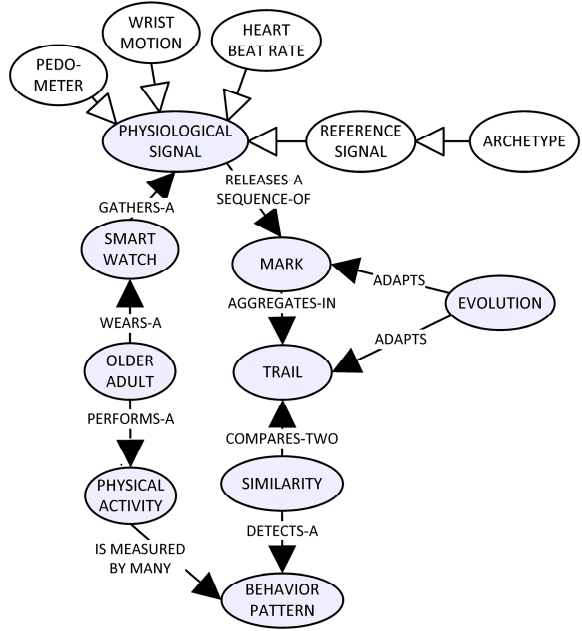

Figure 1: Ontological view of the proposed approach.

More specifically, the behaviour pattern is detected by a computational unit called *Stigmergic Receptive Field* (SRF), shown in Fig. 2. An SRF periodically takes samples of a type of physiological signal. A min-max normalization of the continuous-valued samples is assumed. Normalized samples $d(k)$ feed the clumping process, which is a kind of soft discretization of the samples to a set of parametrized regions of interest. The *clumping* process is implemented by a multi-sigmoidal function, characterized by a couple of inflection points, i.e., $\alpha_I$ and $\beta_i$, for each region of interest. After clumping, each sample $d_C(k)$ enables the release of a mark in a computer-simulated spatial environment. A *mark* is a trapezoid characterized by intensity 1, width $\varepsilon$ and $2\varepsilon$, and position. The position corresponds to the value of the sample $d_C(k)$. Marks $M(k)$ aggregates in the *trail T(k)*, whose intensity is subject to a temporal evaporation. This means that a quantity $\delta$ of $T$ decreases after a step of time. As a consequence, after a certain time an isolated mark disappears, whereas consecutive samples close to a specific region of interest (clump) will superimpose, increasing the trail intensity. In practice, the trail can be considered as a short-term and a short-size action memory.

The Trail captures a coarse spatio-temporal structure in a segment of the domain space (multi-step sliding time window), robust to noise and variability of samples at the micro-level (Avvenuti, 2013). Subsequently, a degree of similarity can be computed comparing two trails generated with different sample streams. At the first level of processing, a segment of the current time series and a segment of an archetype series are compared by means of similarity. The similarity between two trails $T_A$ and $T_B$ is the cardinality of the intersection divided by the cardinality of the union of the trails, i.e., $|T_A \cap T_B| / |T_A \cup T_B|$.

An archetype is a pure form time series which embodies a behavioural class. An example of class in our domain is "Variable-High heartbeat", which means that the heartbeat shows some sudden increases of level over time. Other class provided as a basis of archetypes are *Low*, *Variable-Low*, *Medium*, *Variable-High* and *High*.

After the calculation of similarity, the SRF carries out the *activation*, which increases/decreases the rate of similarity according to a sigmoid with two inflection points. The term activation is taken from neural sciences and it is related to the requirement that a signal must reach a certain level before a processing layer can fire to the next layer.

Each SRF should be properly parameterized to enable an effective samples aggregation and output activation. For example, short-life marks evaporate too fast, preventing aggregation and pattern reinforcement, whereas long-life marks cause early activation.

The Adaptation uses the Differential Evolution (DE) algorithm to adjust the parameters of the SRF (Cimino, 2015), in order to minimize the fitness function, which is computed over a training set of $N$ signals. More in detail, the DE adapts: (i) the clumping inflection points $\alpha_1$, $\beta_1$, $\alpha_2$, $\beta_2$; (ii) the mark width $\varepsilon$ (iii); the trail evaporation $\delta$; (iv) the activation inflection point $\alpha_A$, $\beta_A$. The adopted fitness function is the mean square error, computed as difference between desired and actual output value, evaluated on the training set: $fit = \sum_1^N (s_i - s'_i)^2 / N$. In Fig. 2, $\bar{d}(k)$ and $d(k)$ are the data samples of the reference and current signal, respectively. Both signals periodically feed the SRF, and are processed in parallel up to the similarity, where they are compared.

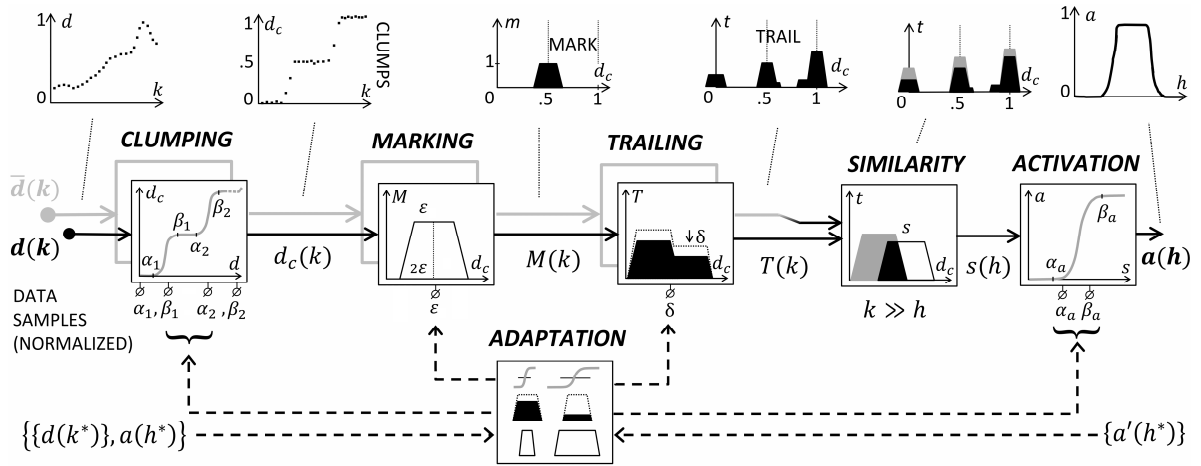

Figure 2: Structure of a Stigmergic Receptive Field.

The modules of the reference signal are represented as grey shadow of the corresponding modules of the current input segment. The training set, on the bottom, is denoted by asterisks: it is a sequence of (input, desired output) pairs, represented on the left. Overall, the SRF plays the micro-pattern detection.

An SRF can be also used in a multi-layered architecture, thus providing further levels of processing to realize a macro analysis. A collection of SRFs specialized on different archetypes is arranged into a connectionist topology, making a *Stigmergic Perceptron* (Fig. 3). The Stigmergic Perceptron detects the similarity between an ordered collection of reference signals and the current input samples, by forming a linear combination of the SRFs with the highest similarity, represented as a circular selector in Fig. 3. The output of each SRF is calculated as a mean of the outputs (represented as 1-to-5 in Fig.3) weighted by the similarities. Each Stigmergic Perceptron is dedicated to a kind of sensor: heart rate, wrist motion and pedometer. Subsequently, the outputs of the three Stigmergic Perceptron are fused via a weighted sum, in order to obtain a combined classification of the effort of each activity segment. Weight are set up via Linear Least Square Method, (Hager, 2012) using a training set made by the multi-sensory input and the expected effort for each type of physical activity.

Finally, the real value representing the current activity segment PAL, it is passed to another SRF aimed to analyze physical activities as a macro-pattern, i.e., the daily PAL. Again, this SRF computes a macro-level similarity between two daily time series: the current and a reference one. An example of class is a *Low PAL Day*, in which user does not perform any intense physical activity. Similarly, a linear combination of similarities among each archetype (*Low*, *Medium* and *High*) represents the daily PAL assessment.

## 4 EXPERIMENTAL STUDIES

To show the effectiveness of the approach, we carried out experiments on three subjects, aged 60, 74, 79, who will be referred to as "A", "B", and "C", respectively. To monitor the PAL of subject A, four weeks of heart-rate, pedometer and accelerometer signals have been gathered, via smartwatch. We ask subject A to keep a diary to annotate begin, end, type and effort of activities performed during each day, as well as the perceived daily effort. After an early visual inspection of the signals, via mouse-based panning in a computer generated figure, and of the corresponding diary entries, the sliding window at the micro-level of pre-processing has been set to 6 minutes. Moreover, a training set of signals which clearly match the physical activities of the diary has been selected.

Since the parameters have a different sensitivity, the adaptation process of the SRFs is made on a two-phase protocol: (i) the global training phase, determining an interval for the evaporation rate of each SRF, which is the most sensitive parameter; more specifically, the interval is determined considering the narrowest interval including the fitness values above the 90th percentile; (ii) the local training phase, made for each SRF separately, by using the interval generated in the first phase; the intervals for the other parameters can be statically assigned on the basis of application domain constraints; the training set for each $i$–th SRF is made by half signals belonging to the $i$–th archetype,

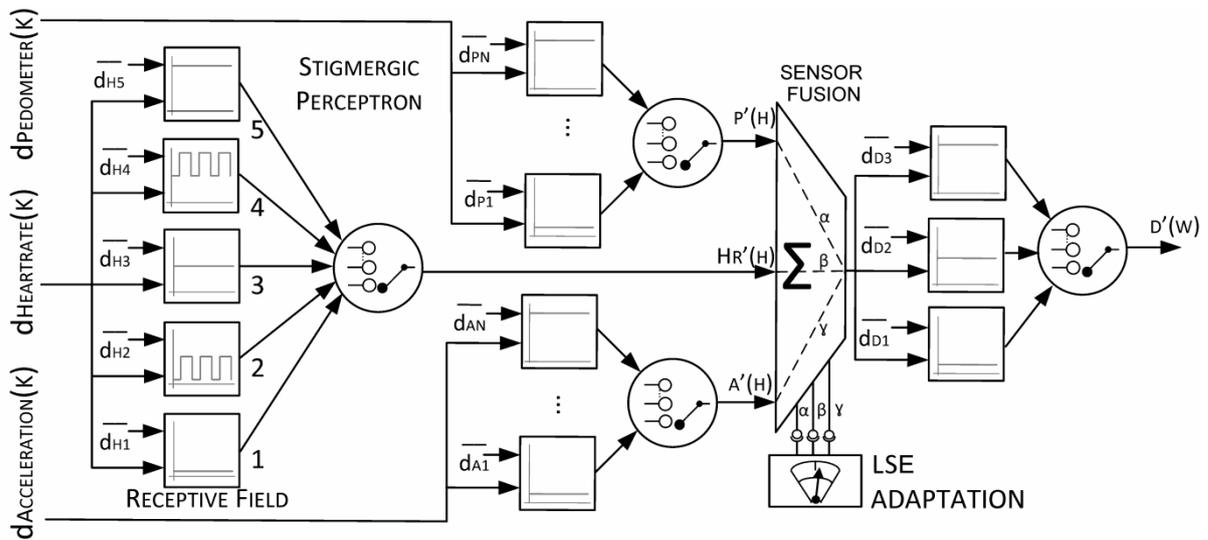

Figure 3: Topology of a multilayer architecture based on Stigmergic Perceptrons.

and half signals belonging to the *i*-1−th and *i*+1−th archetypes.

As target similarity for the fitness function, the values 1 and 0 have been used for similarity and dissimilarity, respectively. As a fitness function, the Mean Square Error is calculated between the similarity computed by the system and the target similarity, for each SRF.

Once the Stigmergic Perceptron have been trained, their outputs, P'(h), HR'(h), and A'(h) in figure, are then provided to the sensor fusion module, which models the mapping from sensor-driven archetypes to the PAL via a linear combination. The weights are determined through the standard Least Square Error optimization, which minimizes the error with respect to the corresponding entries in the diary. After the fusion, a further Stigmergic Perceptron is trained to classify the PAL according to the following archetypes: Low (1), Medium (2), and High (3).

Fig. 4 shows a boxplot of the PALs of Subject A. Here, each input-output pair is calculated over a 6 minute windows, for 165 total windows.

More specifically, Subject A is a healthy and active 60 years man. He works and practises several sports. He does not present any frailty symptom, and is not under drug therapy. His activities data were been collected through smartwatch for a time period of 4 weeks of summer 2016. The activities performed and annotated on the diary spread from walking to excursion, as reported in Fig. 4. The output provided by the system as a PAL is a real number in the interval [1,3], to represent any combination of the classes Low, Medium, High. In Fig. 4, each row comprises the samples related to a specific activity; each column represents a different PAL. On the top of each column, the activities with the expected PAL are also included. In practice, any activity involves a different life cycle with more or less different PALs (e.g. a recover process). In each row, the left and right side of the box represent the first and third quartiles of the distribution, the band inside the box is the second quartile (the median), while the ends of the whiskers represent the maximum and minimum of the distribution.

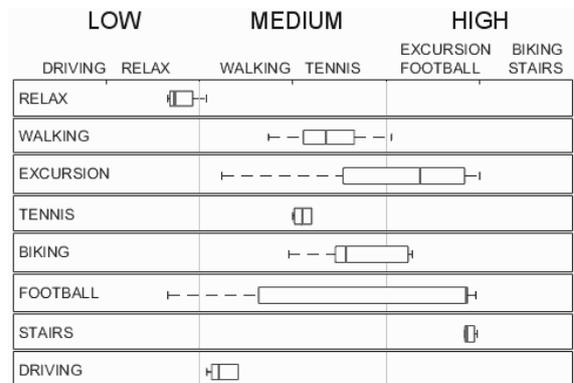

Figure 4: Physical Activity Levels of the Subject A over four weeks.

Overall, the fitting between the expected and the calculated PALs for subject A is good: the Mean Square Deviation over 165 time windows is 0.326. Indeed, we remark that relax, walking, virtual tennis, and stairs activities are mostly included in the expected class. Not surprisingly, excursion and five-a-side football are partially spread on the adjacent class, since the development of this kind of activities involves recovery processes with a lower physical activity level. Similarly, the biking activity is expected to range from medium to high PAL, depending on the speed and the road slope. In contrast driving, which is an activity with constant PAL, is entirely included in the Medium PAL and not, as expected, in the Low. A deep investigation into the levels of processing shows that the most error for driving is located in the sensor fusion. In general, depending on the traffic and anxiety levels, driving may be an activity with high cognitive load, leading to a high heartbeat rate. In addition wrist acceleration is constantly high.

However, driving is not a physical activity. Since wrist motion and heartbeat rate may be both high when driving, the pedometer should play a more important role in decision. For this purpose, the linear combination of features used as the decision function is not a good induction system. We expect that a tree-like structure for the decision function could better distinguish situations in which one sensor can better play the role of discriminator. A comparative study can be proficiently handled as future work.

Since the purpose of the system is to assess physical activity on a daily basis, Fig. 5 shows the daily PAL computed by the system (white circles) along with the expected PAL (black circles). It is computed as the average PAL of the time windows of the day. Here we remark that, in 21 days, there is only one misclassification, on day 17. A deep investigation has shown that the error is derived by the driving activity, which is relevant for day 17. We remark that other 3 days in which driving was not the main activity are not affected by misclassification. Overall, the Mean Square Deviation with respect to the expected daily PAL is 0.158. The system was trained using 9 days (43%) of this data set.

In order to investigate the system behavior on older subjects, we have involved other two subjects into the experimentation. A problem is that older subjects are usually less active and less prone to manage a detailed diary. For this reason, we used the training carried out with the subject A for the initial roll-out of the system on the two subjects. The experimentation was made on three types of activity: relaxing, walking, and stairs climbing, and the diary entries were collected by the observer during direct observation. Although the number of activities and the gathering time are not relevant, results are very promising.

More specifically, subject B is a 74 years old man. He is retired, and is not physically active. He does at most 30 minutes of walking per day, for 5 days per week. He practises gardening, and does not present any frailty symptom. Occasionally he had some fall (recently, when taking the bus) without injuries. He is not under drug therapy. The data, on 14 time windows, were gathered on spring 2016. Each activity effort was classified by subject B as Low for Relax, High for Walking and Stairs climbing. The system performance is measured by a Mean Square Deviation of 0.0533.

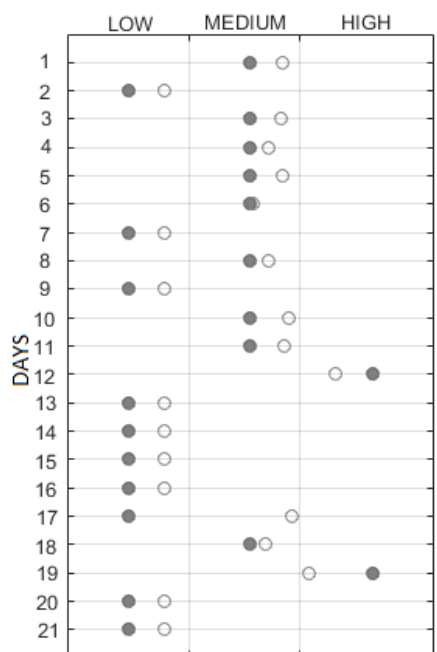

Figure 5: Daily PAL assessing in Subject A.

Subject C is a sedentary 79 years old man. He is retired. He is quite a lazy subject: he walks for less than 15 minutes per day, for 5 days per week. He periodically does medical examinations, and is under drug therapy for blood pressure and cholesterol lowering therapy. The data, on 12 time windows, were gathered on summer 2016. The subject C classified his activity effort as Low for Relax and High for Walking and Stairs Climbing. The system performance is measured by a Mean Square Deviation of 0.0996.

We remark that although both subjects have classified the walking activity effort as high, which is different than subject A, the system has correctly measured the walking. Actually, the direct observation of subjects B and C has clearly shown that walking requires some degree of physical effort for them. The early results show that our system assesses PAL on how the activity is performed, despite of activity type.

# 5 CONCLUSIONS

In this paper an innovative computational architecture for broad-spectrum assessment of the physical activity level of older adults is presented. The detection strategy is founded on stigmergic computing, a bio-inspired mechanism of emergent systems, which requires a continuous data gathering through general-purpose and non-intrusive devices, such as smartwatch. The architectural design is first presented. Then, the system experimentation is discussed on three subjects, making possible the initial roll-out of the approach in real environments. Experimental studies show promising results. A clinical trial could be interesting to validate the approach.

The system performance can be further improved exploiting more sensors and investigating a tree-like structure for the decision function, in order to better distinguish situations in which one sensor plays the role of discriminator.

## ACKNOWLEDGEMENTS

This work was partially supported by the PRA 2016 project "Analysis of Sensory Data: from Traditional Sensors to Social Sensors" funded by the University of Pisa.